# Process stabilization by peak current regulation in reactive high-power impulse magnetron sputtering of hafnium nitride

T. Shimizu,[1,2] M. Villamayor,[1,3] D. Lundin,[4] and U. Helmersson[1]

[1] Plasma & Coatings Physics Division, IFM Materials Physics, Linköping University, Linköping, SE 581-83, Sweden
[2] Division of Intelligent Mechanical Systems, Graduate School of System Design, Tokyo Metropolitan University, 6-6, Asahigaoka, Hino-shi, 191-0065 Tokyo, Japan
[3] National Institute of Physics, University of the Philippines–Diliman, Diliman, Quezon City, 1101 Philippines
[4] Laboratoire de Physique des Gaz et Plasmas - LPGP, UMR 8578 CNRS, Université Paris-Sud, 91405 Orsay Cedex, France

E-mail: simizu-tetuhide@tmu.ac.jp and ulfhe@ifm.liu.se

**Abstract.** A simple and cost effective approach to stabilize the sputtering process in the transition zone during reactive high-power impulse magnetron sputtering (HiPIMS) is proposed. The method is based on real-time monitoring and control of the discharge current waveforms. To stabilize the process conditions at a given set point, a feedback control system was implemented that automatically regulates the pulse frequency, and thereby the average sputtering power, to maintain a constant maximum discharge current. In the present study, the variation of the pulse current waveforms over a wide range of reactive gas flows and pulse frequencies during a reactive HiPIMS process of Hf-N in an Ar–$N_2$ atmosphere illustrates that the discharge current waveform is a an excellent indicator of the process conditions. Activating the reactive HiPIMS peak current regulation, stable process conditions were maintained when varying the $N_2$ flow from 2.1 to 3.5 sccm by an automatic adjustment of the pulse frequency from 600 Hz to 1150 Hz and consequently an increase of the average power from 110 to 270 W. Hf–N films deposited using peak current regulation exhibited a stable stoichiometry, a nearly constant power-normalized deposition rate, and a polycrystalline cubic phase Hf-N with (111)- preferred orientation over the entire reactive gas flow range investigated. The physical reasons for the change in the current pulse waveform for different process conditions are discussed in some detail.

Keywords: reactive sputtering, HiPIMS, hafnium nitride, process control

# 1. Introduction

Reactive magnetron sputtering is a physical vapor deposition technique which is applied to grow compound films such as nitrides, oxides, and carbides. An inherent feature in reactive magnetron sputtering is process instability due to the complex relation between the fluxes of reactive gas and sputtered metal from the cathode [1]. High reactive gas fluxes induce compound formation on the cathode (target), which is referred to as target poisoning. Since the sputtering efficiency of the compound material is typically lower than that of the corresponding metal target, poisoning usually causes a significant decrease in flux of metal from the cathode[1]. Therefore, to obtain stoichiometric compound films with relatively high deposition rates, the process generally has to be maintained in the transition region between the metallic and the compound mode, where instability due to hysteresis is commonly observed [2].

A number of studies have been performed to stabilize the reactive process in the transition regime and to obtain stoichiometric deposited layers, for example, by increasing the pumping speed [3], reducing the target area [4] or by utilizing a fast feedback control of the reactive gas flow by monitoring the partial pressure [5], by monitoring the optical emission of the plasma, which is called plasma emission monitoring (PEM) [6], or by varying the cathode voltage [7]. Some of the drawbacks from these feedback control systems of the reactive gas flow are the additional complexity and cost for the monitoring devices and the reaction time delay in gas flow control to stabilize the plasma state. Thus, a more simple approach to stabilize the operating point in reactive sputtering without any additional monitoring devices and with a shorter time constant for modifying the control parameters is desired.

Meanwhile, high power impulse magnetron sputtering (HiPIMS) has received much attention as an alternative to conventional direct current (DC) sputtering during the last decade [8]. In HiPIMS, a high voltage is applied to the target over short pulses with a low duty cycle of less than 10% at a low frequency of less than 10 kHz, resulting in extremely high peak target power densities of several kWcm$^{-2}$ [9]. The main principles of the HiPIMS process is described in detail elsewhere [10]. A recent study on reactive HiPIMS shows that the hysteresis can be reduced or completely eliminated [11] for a selected range of process parameters [12]. This characteristic of HiPIMS offers a great potential to stabilize the deposition process by controlling the growth of the poisoned layer on the target surface. However, since this property is strongly dependent on the system itself as well as the process conditions, process stabilization in reactive HiPIMS is still needed.

Several approaches have been used to minimize the instabilities and to enhance the deposition rate, e.g. a PEM based gas flow control technique [13] or a pulsed reactive gas flow control [14]. Instead of using the flow control, Sittinger *et al.* [15] proposed adjusting the average power by altering the pulse frequency to stabilize the set point of the reactive gas partial pressure. By using the partial pressure as a reference value for the frequency control loop, deposition of aluminum-doped zinc oxide films was successfully stabilized in the transition regime. In a similar way, Weichart *et al.* [16] focused on the variation of discharge current depending on the state of the sputter target composition for sputtering of Al in an Ar/O$_2$ mixture. In the poisoned mode, the discharge current increased beyond the maximum current values attained in the metal mode. This was believed to be related to the higher secondary electron emission coefficient for oxidized aluminum relative to that for metallic aluminum [16]. However, Aiempanakit *et al.* [17] stated that the large discharge current in oxide mode are mainly attributed to the large ionic current due to the impingement of a high fraction of O$^{1+}$ ions. Hala *et al.* [18] have shown a peak current dependency on the pulse frequency, in which a higher peak current was recorded at lower pulse frequency. This was explained by the effect of available time for target oxidation in between the pulses by changing the frequency. Similar results have been reported by Ganesan *et al.* [19] for sputtering of Hf in an Ar/O$_2$ mixture. They demonstrated that the degree of oxidation on the target decreased for longer pulses. As a consequence, the target current decreases due to the decrease of ion-induced secondary electron emission, $\gamma_{SE}$. These variations in the discharge current behaviour in the reactive sputtering process have been observed for different target materials, e.g. Nb [18], Hf [19], Ti [20], Si, Ta [21], Ru [22] in Ar/O$_2$ gas mixture, and for Cr [23] and graphite [24] in Ar/N$_2$ mixture and for graphite in Ar/CF$_4$ atmosphere [25]. As can be clearly seen from these reports, the discharge current is promising reference parameter to monitor the target surface composition in reactive sputtering. By using this reference value as an input to the process control, a desired film stoichiometry can also be achieved in the unstable process region.

The objective of the present study is therefore to explore the feasibility of a peak target current regulation technology in reactive HiPIMS to stabilize the transition zone during reactive deposition over a wide range of experimental conditions. As a simple and cost effective approach to stabilize the discharge current at a given set point, the present study implements a feedback control system by automatic regulation of the pulse frequency. To this purpose, we first monitor waveforms of the real-time pulse current to study the process characteristics during reactive HiPIMS of Hf in different Ar/N$_2$ gas mixtures and for different pulse frequencies. Based on this characterization, the possibility of stabilizing the reactive HiPIMS process by the discharge current characteristics under the control of the pulse frequency is investigated. Finally Hf–N films were grown under peak current, $I_{pk}$, regulation and their stoichiometric properties, deposition rates and crystal structures were evaluated and compared to that of conventional reactive HiPIMS without $I_{pk}$ regulation.

## 2. Experimental details

Experiments were performed in an ultra-high vacuum stainless-steel chamber with a base pressure below $10^{-6}$ Pa. A Hf (99.9 % purity) disk with a diameter of 76.2 mm and a thickness of 6.35 mm was used as sputtering target. Ar gas with a purity of 99.9997% was introduced into the chamber through a mass flow controller and the Ar flow was adjusted to maintain a constant partial pressure of 0.4 Pa. N$_2$ gas (99.9995% in purity) was introduced into the chamber and the N$_2$ flow rate was varied from 2.0 to 4.0 sccm to grow films in the transition, and the compound sputtering modes.

Average sputtering power delivered to the target was varied in the range from 100 to 200 W depending on the discharge current for a constant applied voltage of 450 V. Unipolar pulses with a length of 25 μs and a frequency in the range from 450 to 1200 Hz were supplied by a HiPSTER 1 pulsing unit (Ionautics AB) fed by an MDX 1 K DC power supply (Advanced Energy). The HiPSTER unit was modified to realize PID-regulation of the frequency based on the internally recorded $I_{pk}$ value. The discharge current and voltage time characteristics were recorded by the HiPSTER 1 and monitored on a Tektronix TDS 520 C digital oscilloscope directly connected to the pulsing unit. The power was subsequently obtained by the product of the voltage and the current signals. Measurements of the time-dependent target voltage revealed nearly rectangular waveforms. For the purpose of process stability at various deposition conditions, the target (discharge) current was recorded as a function of the $N_2$ flow and pulse frequency. To study the effects of the $I_{pk}$ regulation by pulse frequency control in the actual reactive deposition process, Hf–N films were deposited onto Si (100) substrates. Prior to the deposition, the Si substrates were cleaned ultrasonically in acetone and isopropanol. Film uniformity was assured by substrate rotation with the speed of 10 rpm. Hf-N films were deposited for 1 hour/sample. Film thicknesses were determined by cross section scanning electron microscopy (LEO 1550 Gemini) and the deposition rates were calculated based on these values. The effect of the deposition conditions on the crystal structure was investigated by means of grazing incidence X-ray diffraction (GI-XRD). The GI-XRD measurements were performed using a PANalytical X'pert diffractometer mounted with a hybrid monochromator/mirror, operated at 40 kV and 40 mA with a Cu anode (Cu Kα, λ=1.540597 Å). The incident beam angle, $\omega$, was 4° while the scanning range in $2\theta$ was 20–80°.

## 3. Results

*3.1. Characterization of the reactive process*

Selected current waveforms at various $N_2$ flow rates in the reactive Hf HiPIMS process are plotted in figure 1. The Ar partial pressure was kept constant at 0.4 Pa in all experiments and the $N_2$ flow was increased from 1.5 to 4 sccm in a controlled manner. A constant pulse voltage of 450 V was applied at the generator output. The pulse conditions were fixed to a duration of 25 μs and a frequency of 600 Hz. The average target power varied in the range of 100 to 140 W due to the change in pulse current. As can be seen from the variation of the waveforms, the slope at the initial rise of the waveforms gradually becomes less steep with increasing N2 flow from 1.5 to 1.9 sccm. The $I_{pk}$ value starts to rise at 2.1 sccm and it increases almost linearly from 2.1 to 4 sccm. A similar trend with increasing reactive gas flow were observed in other studies as mentioned [18-25].

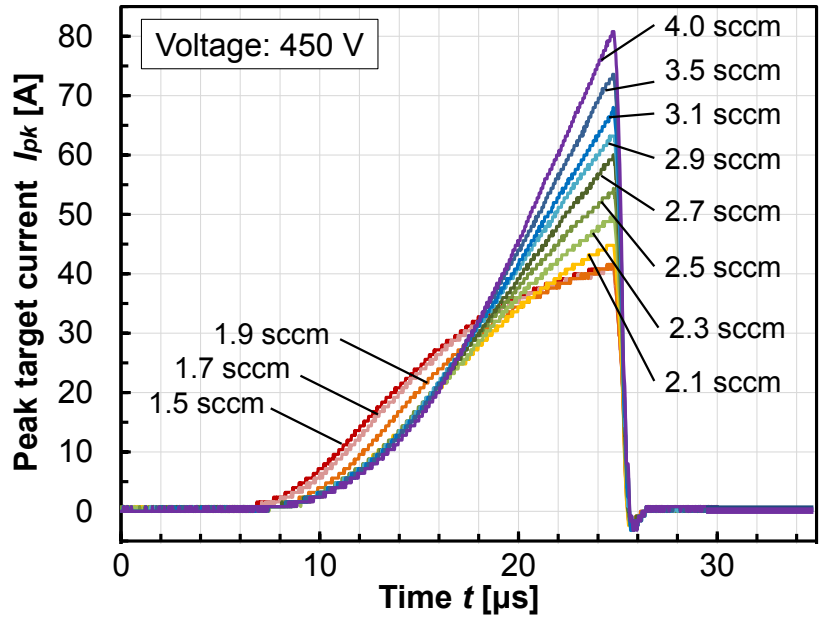

**Figure 1.** The effect of increasing nitrogen flow on the pulse current waveforms of the HiPIMS discharge.

Figure 2 presents the variation of the peak target current (a) and the average discharge power (b) for both increasing and decreasing $N_2$ flow rates. It can be seen that both the target current and the corresponding average power value have a narrow hysteresis between the increasing and decreasing $N_2$ flow. Since there is no abrupt variation at a certain critical $N_2$ flow, this hysteresis might not be a steady state hysteresis, but possibly be due to time-dependent dynamic behaviour, such as an effect of reactive gas gettering at the target surface and chamber walls, which is strongly depending on the process history [26, 27]. When the $N_2$ flow is increased, the peak target current exhibits a gradual increase at $N_2$ flow value of about 2.0 sccm indicating the transition from the metallic to the compound sputtering mode. Under higher $N_2$ flow of more than 3.0 sccm, the target is likely fully covered by a compound nitride layer, as can be seen from the linear increase and decrease in target current. The transition from the compound back to the metallic mode (when the $N_2$ flow is decreased) takes place at an $N_2$ flow of about 2.5 sccm. Similarly to the variation of peak target current, the average target power also exhibited the corresponding hysteresis and the transition regime in the range of $N_2$ flow of 2.0 – 2.5 sccm. The trend with a local minimum at a $N_2$ flow of 1.9 sccm may indicate the initiation of the formation of a nitride layer on the target. However, from the view of the $I_{pk}$ control by frequency, as intended in the present study, the process cannot be controlled at this local minimum, and thus it is here used as a lower limit of our selected process window.

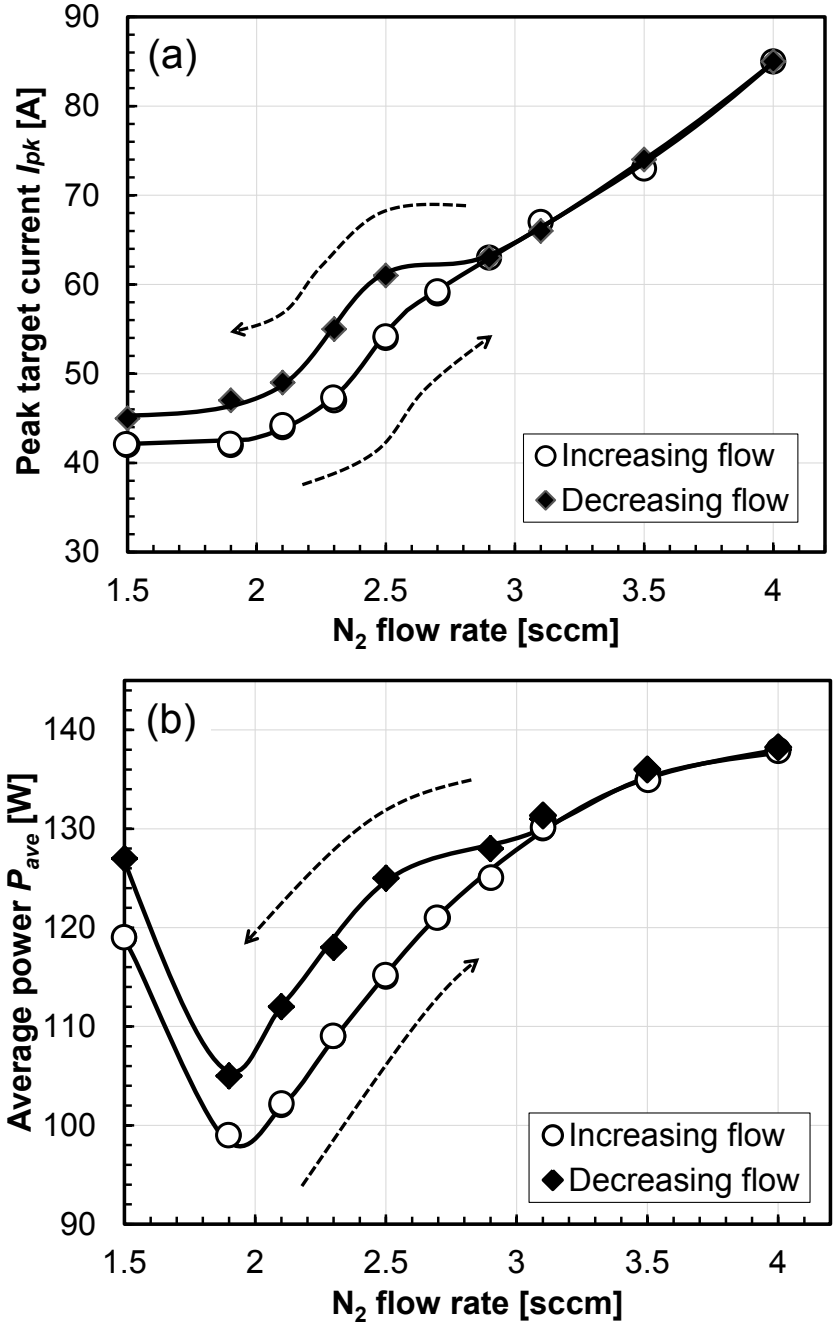

**Figure 2.** Variations of (a) $I_{pk}$ and (b) average power as a function of $N_2$ gas flow rate. Lines are guide for the eyes.

To study the dependence of the $I_{pk}$ versus the pulse frequency, current waveforms at several pulse frequencies ranging from 450 - 1000 Hz were obtained as shown in figure 3. A discharge voltage of 450 V, a pulse duration of 25 μs and a $N_2$ flow of 2.3 sccm were used throughout the experiment. As similarly observed in the trend in the variation of gas flow, the onset of the current waveform becomes slower with decreasing pulse frequencies. Additionally, the peak current value simultaneously increases and the lowest frequency shows the highest peak current.

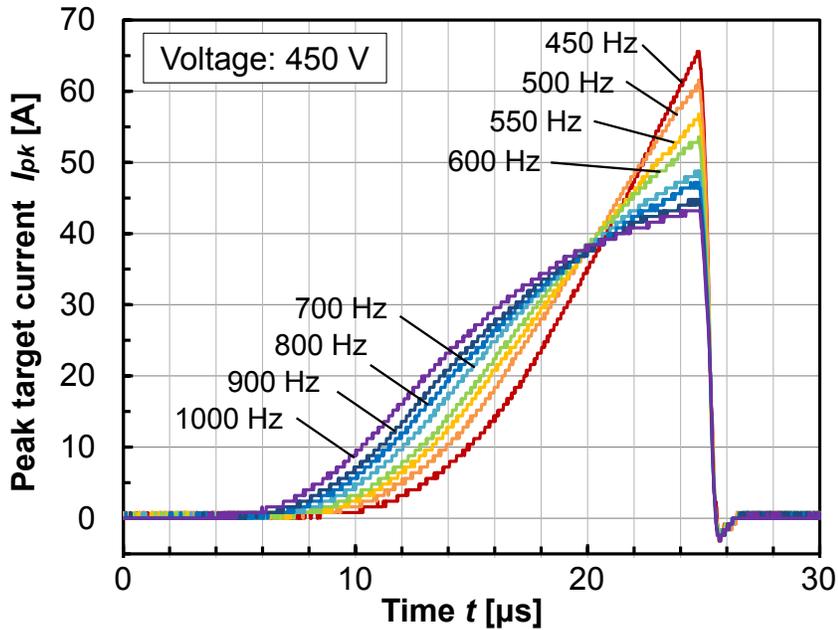

**Figure 3.** The effect of the pulsing frequency on the pulse current waveforms of the HiPIMS discharge.

Figure 4 summarizes this variation of peak target current depending on the pulse frequency and corresponding average power. As can be expected, the average target power increases with increasing pulse frequency and as a consequence $I_{pk}$ decreases, since the compound layer of the target surface is reduced by the increased sputtering as the target enter the metallic mode. In contrast, at the lower average power under lower pulse frequency the current pulses (figure 3) resembled the ones found in the poisoned mode in figure 1, characterized by less target sputtering. Consequently, it is shown that the average power was successfully altered by controlling the frequency to change the state of target surface, thus the peak target current varies.

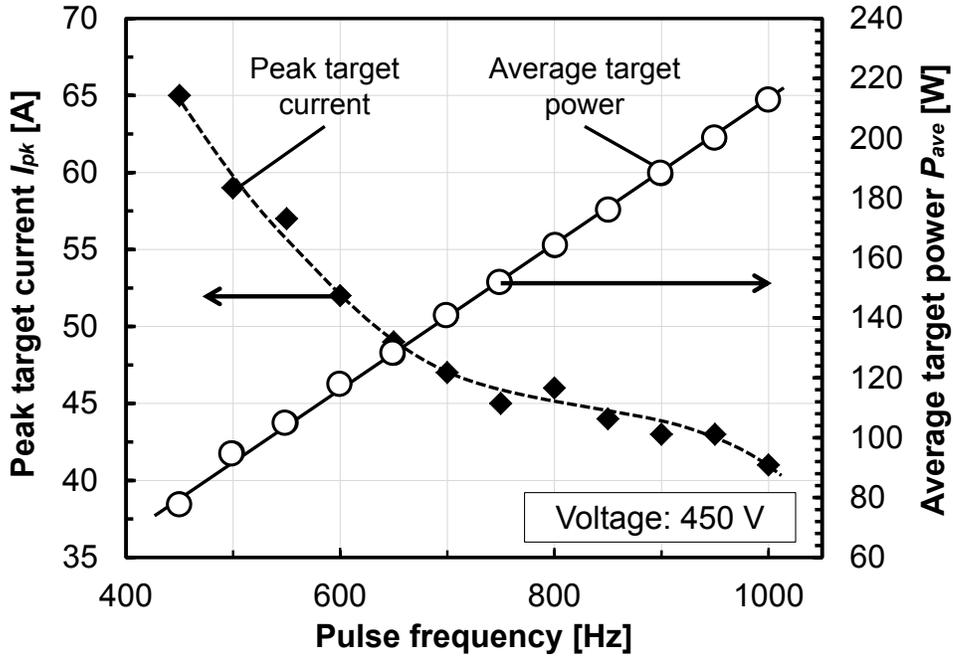

**Figure 4.** Variations of average power and $I_{pk}$ as a function of pulse frequency. Lines are guides for the eye.

*3.2. Peak target regulation by pulse frequency control*

A new control system was built into the HiPIMS pulser to investigate the possibility of stabilizing the reactive HiPIMS process in the transition regime based on the discharge current characteristics. The main feature consisted of a feedback control loop of the real-time pulse current to keep the peak target current at a given set point by automatic regulation of the pulse frequency. According to the trend of the $I_{pk}$ variation as shown figure 2, a peak target current of $I_{pk}$ = 50 A was chosen as a reference value to stabilize the process in the transition regime and it was kept constant by controlling the pulse frequency while varying the N$_2$ gas flow from 2.1 to 3.5 sccm. Figure 5 (a) shows the pulse current waveforms versus N$_2$ flow under $I_{pk}$ regulation mode. The discharge voltage was kept constant at 450 V for all conditions. The current waveforms of the process without $I_{pk}$ regulation are also depicted as a reference, figure 5 (b). By controlling the pulse frequency to keep the $I_{pk}$ value constant, almost the same current waveforms were obtained for different N$_2$ gas flows. For example, for a N$_2$ gas flow of 3.5 sccm, the discharge current curve peaking at 74 A in figure 5 (b) could be reduced and actively stabilized to the set peak value of 50 A (figure 5 (a)) by altering the frequency from the previously set value of 600 to 1150 Hz. By this frequency increase, the average power increased from 135 to 270 W leading to an increased sputtering of the compound layer.

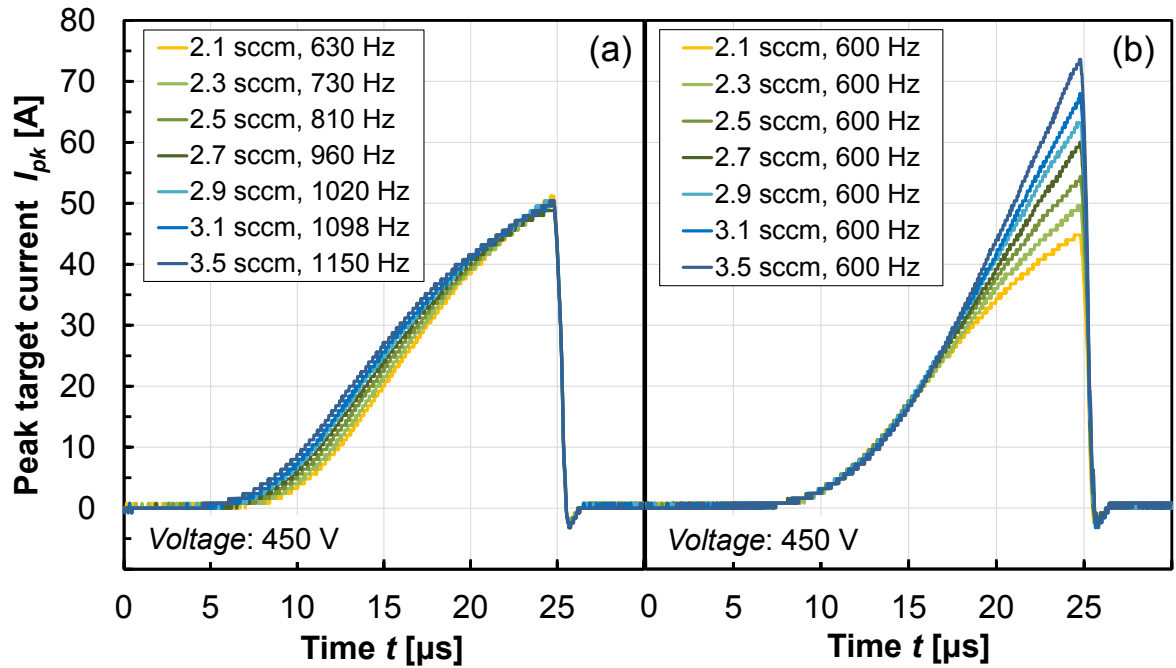

**Figure 5.** Comparison of the pulse current waveforms (a) with and (b) without $I_{pk}$ regulation by pulse frequency control

Variation of the feedback controlled pulse frequencies using the $I_{pk}$ regulation with increasing and decreasing $N_2$ gas flow are summarized in figure 6. For this large variation of the $N_2$ flow rates from 2.1 to 3.5 sccm, the pulse frequency automatically adjusted in the range of 600 to 1150 Hz to keep $I_{pk}$ constant. As can be seen in the figure the frequency increased almost linearly with increasing $N_2$ flow. Also the corresponding average power values (not shown) follow a linear trend from 110 to 270 W.

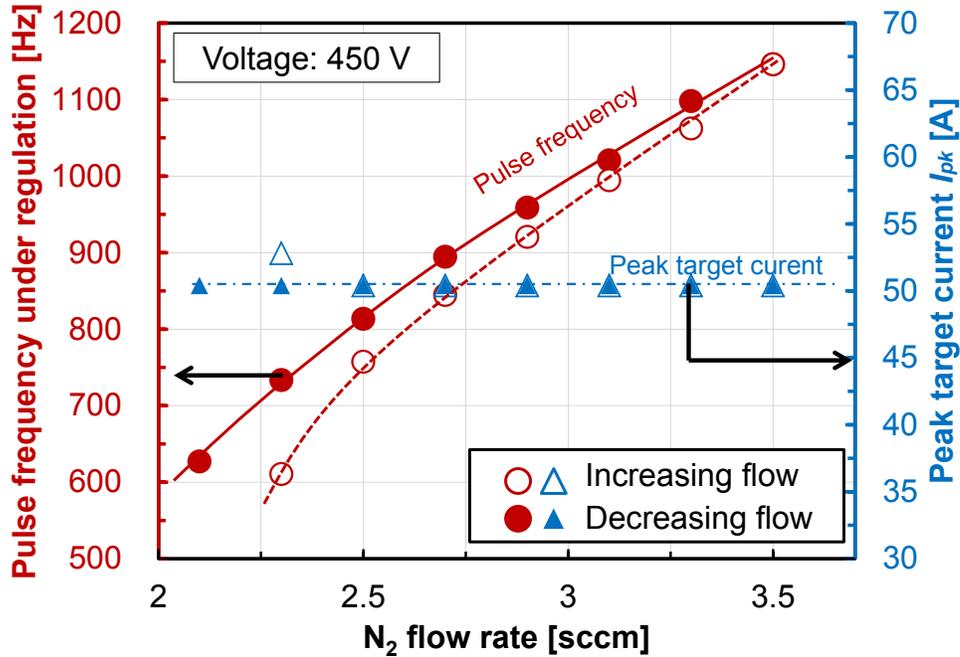

**Figure 6.** Response of controlled pulse frequency using $I_{pk}$ regulation versus $N_2$ gas flow rate. Lines are guides for the eye.

*3.3. HfN deposition under peak current regulation*

The applicability of the $I_{pk}$ regulation by the above described feedback control loop on the actual film growth of HfN was evaluated by comparing the film properties of Hf–N films deposited at various $N_2$ gas flows of 2.0, 2.4, 2.8, 3.2, and 3.6 sccm with and without $I_{pk}$ regulation. Both processes were operated at constant pulsing conditions (450 V and 25 μs pulse width). The trends of the variation of current waveforms with increasing gas flow and the controllability of the peak target current by pulse frequency showed good agreement with the results presented in figures 5 and 6.

Approximate stoichiometries were estimated by visual colour variation of the deposited films. Variations of the colour by altering the $N_2$ flow were clearly observed from the films deposited without $I_{pk}$ regulation. As can be estimated from the gold-like yellow colour of the deposited films, stoichiometric HfN films were obtained for a $N_2$ flow of 2.4 sccm, while the film deposited at the highest $N_2$ flow of 3.6 sccm showed the over-stoichiometric colour of dark brownish or purple [28]. On the other hand, the films deposited with $I_{pk}$ regulation exhibited a much broader process window. In this case, all deposited films at different $N_2$ flow rates (2.0-3.6 sccm) had visually the same golden colour of the stoichiometric composition.

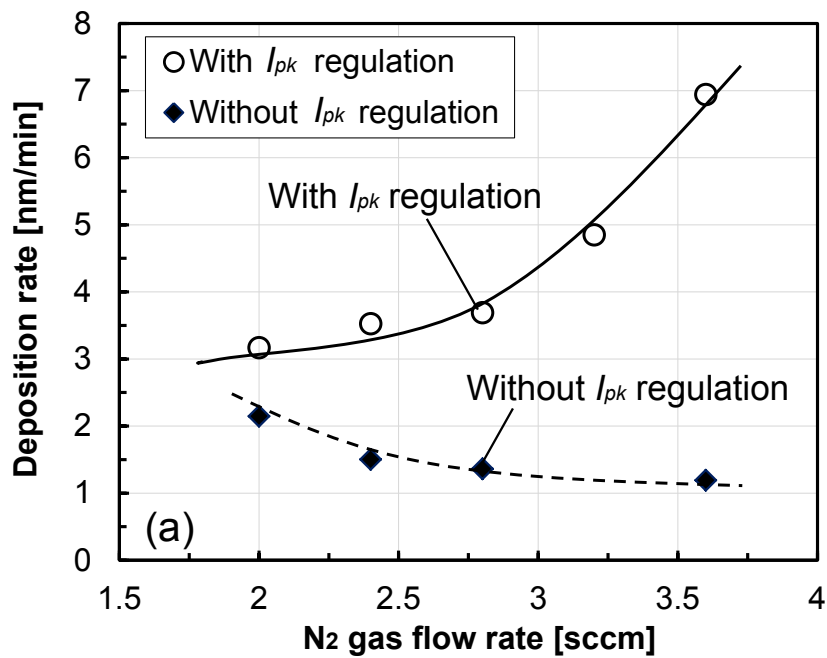

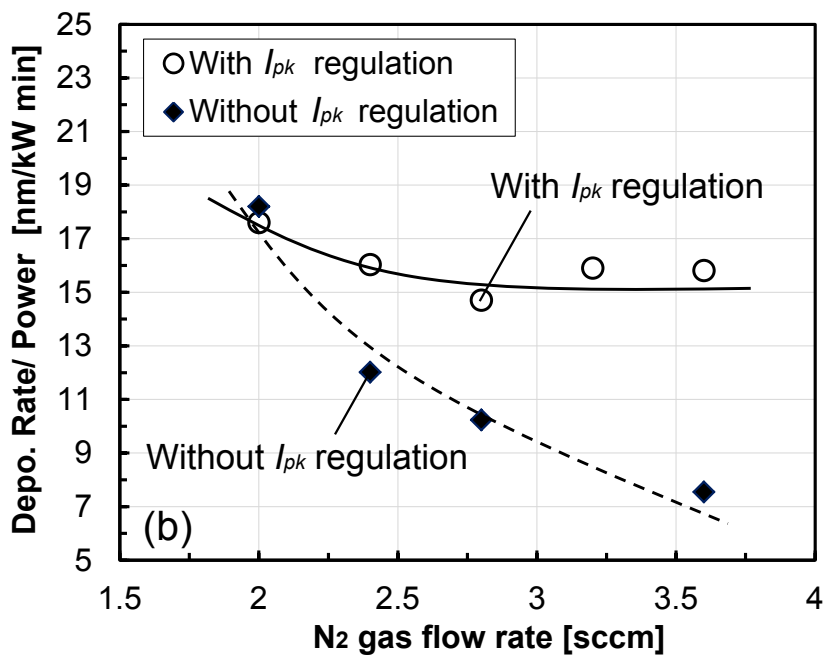

**Figure 7.** Comparison of (a) the deposition rate and (b) power-normalized deposition rate of HfN films deposited with and without $I_{pk}$ regulation by pulse frequency control. Lines are guides for the eye.

The deposition rates with and without $I_{pk}$ regulation are shown in figure 7 (a) as a function of N$_2$ flow rate. The films deposited with $I_{pk}$ regulation demonstrate a rapid increase of deposition rate with increasing N$_2$ flow, while the deposition rate without regulation decreases as a consequence of target poisoning. The result for regulation is clearly due to the relatively higher applied average target power by increasing the pulse frequency for the higher $N_2$ flow conditions. In order to compare the deposition process efficiency, the deposition rates were normalized by the applied power as shown in figure 7 (b). While the power-normalized deposition rate again decreases for the process without $I_{pk}$ regulation as expected, the value stays almost constant using the $I_{pk}$ regulation.

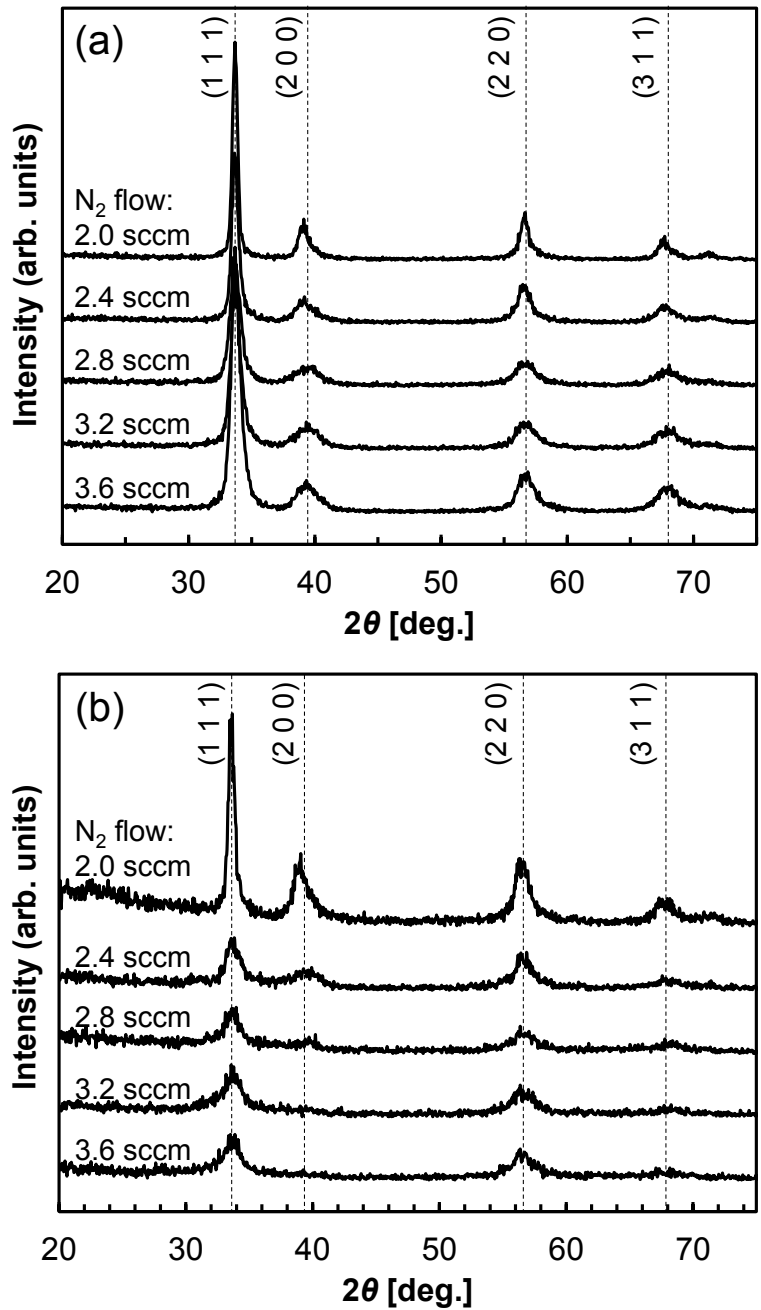

**Figure 8.** GI-XRD $2\theta$ scans obtained from Hf-N films grown at various $N_2$ flow rates ranging from 2.0 to 3.6 sccm (a) using $I_{pk}$ regulation (b) without $I_{pk}$ regulation.

A series of GI-XRD scans of the Hf-N films deposited with $I_{pk}$ regulation at various $N_2$ flows from 2 to 3.6 sccm are shown in figure 8 (a). Scans obtained from all the films grown under $I_{pk}$ regulation contain (111), (200), (220), and (311) diffraction peaks from polycrystalline NaCl-structure Hf-N [29]. The diffraction peak positions were located fairly close to the expected diffraction angles with respect to reference HfN powder patterns [30] (e.g., $2\theta_{111}$= 33.65° and $2\theta_{200}$= 39.20° for N$_2$ flow of 2.4 sccm vs. $2\theta_{111}$= 34.10° and $2\theta_{200}$= 39.58° for HfN powder diffraction patterns, respectively.) Irrespective of the N$_2$ flows provided, sharp diffraction peaks with a preferred (111) orientation were observed. This result was obtained from integrated diffraction peak intensities $I_{hkl}$ which are normalized to corresponding results from powder diffraction patterns. Additionally, the relative peak intensities were almost identical for all the different N$_2$ flow rates investigated. The films deposited without $I_{pk}$ regulation, as shown in figure 8 (b), exhibit different diffraction peaks depending on the N$_2$ flows. A diffraction pattern with preferred (111) orientation, as obtained under $I_{pk}$ regulation, is in the non-regulated case only found for a N$_2$ flow of 2.0 sccm. The (111) peak is diminished by increasing the N$_2$ flow. These variations in the diffraction patterns depending on the N$_2$ flows might be explained by a distortion of the cubic symmetry due to the interstitial incorporation of the increasing contents of nitrogen [31]. However, an in-depth discussion of the crystal phase and texture of Hf-N films is beyond the scope of the present study. Still, the variations in the crystal structure due to the different N$_2$ flows are stabilized by regulating the $I_{pk}$ value using pulse frequency control.

## 4. Discussion

As demonstrated above, the reactive Hf HiPIMS process in Ar/N$_2$ mixtures can be stabilized by regulating the frequency to maintain a constant peak value of the current in the pulse, the $I_{pk}$. This is likely to work for all process conditions where there is a simple relation between $I_{pk}$ and the reactive gas content in the process chamber. The actual reactive gas content was not measured in this work, but since there is only a minor hysteresis observed in the present case, the gas nitrogen content scales at least with a simple relation to the N$_2$ flow for flows higher than 2 sccm (see figure 2(b)) where $I_{pk}$ increases with flow. For the lower flows, the $I_{pk}$ increases again for decreasing flows. Therefore, for the present approach using $I_{pk}$ as control parameter, process conditions are limited to N$_2$ flow rates above 2 sccm. There are of course other possible control parameters associated with the pulse shape that can be used, such as the average peak power or the derivative of the discharge current. We have not yet explore this option this and the optimal control parameter is likely to be varying with material system used or even with system configurations used, such as the magnetic arrangement of the magnetron source. We also foresee that analyses of the current pulse form can be used to probe the degree of target erosion.

In view of the observed results it is interesting to discuss the physical reasons for the change in the current pulse shape with change in process conditions. The observed increase in $I_{pk}$ with increasing N$_2$ flows in the later stage of the HiPIMS discharge pulse, as shown in figure 1, is likely a consequence of a higher $N^{1+}$ fraction in the target vicinity that is taking part in the sputtering process, in line with what has been reported by Magnus *et al.*

[32] for sputtering Ti in Ar/N$_2$. Large fluxes of N$^{1+}$ to the cathode will contribute to a strong increase of the discharge current carried by ions. In the present case, the density of N$^{1+}$ will likely not decrease as compared to the partial pressure of N$_2$, which undergoes gas rarefaction, due to the higher partial sputtering in combination with a continuous supply of N atoms from the nitrided target (as similarly observed in Ar/O$_2$ gas mixture [17]). In addition, in comparison with the Ti target, gas rarefaction [33] in the present Hf-N process becomes more prominent by transferring larger momentum to Ar in front of the target, due to a larger collision cross-section and due to a higher mass of Hf as compared to Ti [34]. The overall consequence is an increased probability of self-sputtering, commonly seen in HiPIMS pulses [35], by the Hf ions and/or N$^{1+}$ ions as the discharge current increases [36]. Furthermore, the secondary electron emission increases in the self-sputtering regime when sputtering the target with N$^{1+}$ instead of Hf ions resulting in an increased discharge current. This can be understood when considering that the secondary electron emission can only occur if $0.78\varepsilon_{iz} > 2\phi$, where $\varepsilon_{iz}$ is the neutralization energy of the ion and $\phi$ is the work function of the target surface [37]. In the case of Hf$^{1+}$ ions with $\varepsilon_{iz} = 6.65$ and $\phi_{HfN} = 4.65$ [38] this criterion is not fulfilled and the secondary emission yield $\gamma_{SE} \approx 0$, wheras for N$^{1+}$ ions $\varepsilon_{iz} = 14.5$, which leads to $\gamma_{SE} > 0$.

Another observation is the initial slower current rise seen for larger N$_2$ flows (figure 1). Since this takes place in the very early phase of the discharge pulse, which is dominated by Ar ion bombardment [17], it is unlikely that it is an effect of self-sputtering as above. Instead, we believe that it is due to a change of the $\gamma_{SE}$ and the sputter yield $Y$ as a compound layer is formed on the surface of the Hf target. No values of the $\gamma_{SE}$ for the Hf-N layer has been found in the literature, but commonly a decrease is seen for increasing N-coverage in most metal targets sputtered in an Ar-N$_2$ mixture [39]. The lower $\gamma_{SE}$ in the compound mode can prolong the buildup of charge carriers in the bulk plasma, which leads to a lower discharge current being drawn during the initial current rise. In addition, a decrease of sputtered material with increasing N$_2$ flow rate has been found in the present work as shown in figure 8 in Section 3.3. This also contributes to a slower formation of ions from the sputtered material and thereby a reduced flux of metal ions being back-attracted to the target carrying the discharge current [17].

A similar tendency of a small delay in the onset of the discharge current was also seen in the pulse current waveforms under $I_{pk}$ regulation as shown in figure 5 (a). One possible explanation for this small delay might be the faster current rise due to the effect of seed charge from the previous pulse under higher pulse frequency. In fact, Barker *et al.* [40] showed in their approach with chopped HiPIMS of Ti in an Ar atmosphere that there is a faster pulse current rise for shorter pulse off-times. Thus, seed charge can remain from the previous pulse and it contributes to the faster initiation of the discharge pulse current. However, since the pulse off-time provided in the present study is more than 1500 µs, the effect of this seed charge from the previous pulse might be small. Barker *et al.*, showed that the effect of the pulse off-time becomes small for off-times longer than 250 µs. An alternative explanation can be an increased $\gamma_{SE}$ when the target is exposed for a shorter time to nitrogen atmosphere in-between the pulses, as similarly explained above. This means that the current curves recorded at a higher N$_2$ flow rate and higher pulse frequency in figure 5 (a) are likely characterized by a smaller fraction of compound layer on the target and thereby a higher $\gamma_{SE}$, which speeds up the buildup of charge carriers in the bulk plasma and leads to a higher discharge current being drawn during the initial current rise. This trend in the onset of the discharge current depending on the pulse frequency is also clearly observed from figure 3. However, no such changes are seen in figure 5(b), where the $I_{pk}$ regulation is not used. In this figure the frequency is fixed at 600 Hz and the 2.1 sccm case is already in the transition zone, and hence characterized by a low $\gamma_{SE}$, which is also shown in figures 1-2.

## 5. Conclusion

In the present study, the variation of the pulse current waveforms over a wide range of reactive gas flows and pulse frequencies during a reactive HiPIMS process of Hf-N in an Ar–$N_2$ atmosphere have been investigated. The simple relation between the peak target current and the $N_2$ flow (for flows higher than 2 sccm), suggests that the discharge current can be used as an effective reference parameter to stabilize the reactive process at any given set point in the transition regime. Based on this characterization, a feedback control loop of the peak target current by the automatic regulation of the pulse frequency was implemented. Accurate system response to the variation in target surface state from compound to metallic mode was demonstrated. Furthermore, Hf–N films were deposited using the $I_{pk}$ regulation mode. A wider process window of reactive gas flow compared to the conventional HiPIMS process without $I_{pk}$ regulation was demonstrated with maintained stoichiometric composition, a nearly constant power-normalized deposition rate, and a polycrystalline cubic phase Hf-N with (111)-preferred orientation. Consequently, a simple and cost effective control technique based on $I_{pk}$ regulation shows great potential to provide precise control and stable operation of reactive HIPIMS discharges.

**Acknowledgements**

This work has been supported by the Swedish Research Council (VR) through contract (Contract No. VR 621-2014-4882). One of the authors Shimizu acknowledges the funding from Japan Society for the Promotion of Science (JSPS), in the form of a Grant-in-Aid for Young Scientists B (No.26820327) and from the AMADA foundation for a Grant-in Aid for general research and development (AF-2013028). Villamayor would like to acknowledge the Bases Conversion Development Authority (BCDA), Philippines for funding the Department of Science and Technology (DOST) "Bridging the Human Competency Gaps in Support for the National R & D Agenda" Post-doctoral Fellowship grant. We would also like to thank Petter Larsson at Ionautics AB for technical support.